\documentclass[prd,amsmath,amssymb,superscriptaddress,12pt]{revtex4-1}

\usepackage{color, colortbl}
\usepackage{graphicx}% Include figure files
\usepackage{dcolumn}% Align table columns on decimal point
\usepackage{mathtools}
\usepackage{bm}
\newcommand{\beq}{\begin{equation}}
\newcommand{\eeq}{\end{equation}}
\newcommand{\ba}{\begin{array}{ccc}}
\newcommand{\ea}{\end{array}}

\def\bea{\begin{eqnarray}}
\def\eea{\end{eqnarray}}
\usepackage{color}
\usepackage{float}

\begin{document}
\title{Machine learning phases of matter} 

\author{Juan Carrasquilla}
\affiliation{Perimeter Institute for Theoretical Physics, Waterloo, Ontario N2L 2Y5, Canada}
\author{Roger G. Melko}  
\affiliation{Department of Physics and Astronomy, University of Waterloo, Ontario, N2L 3G1, Canada}
\affiliation{Perimeter Institute for Theoretical Physics, Waterloo, Ontario N2L 2Y5, Canada}

%\date{\today\\
%\vspace{-1.6in}}

\begin{abstract}

Neural networks can be used to identify phases and phase transitions in condensed matter systems via supervised machine learning.
Readily programmable through modern software libraries, we show that a standard feed-forward neural network can be trained to detect multiple types of order parameter directly from raw state configurations sampled with Monte Carlo.
In addition, they can detect highly non-trivial states such as Coulomb phases, and if modified to a convolutional neural network, topological phases with no conventional order parameter.
We show that this classification occurs within the neural network without knowledge of the Hamiltonian or even the general locality of interactions.
These results demonstrate the power of machine learning as a basic research tool in the field of condensed matter and statistical physics.
\end{abstract}
%\pacs{}
\maketitle
Condensed matter physics is the study of the collective behavior of massively complex assemblies of 
electrons, nuclei, magnetic moments, atoms or qubits~\cite{wen2004quantum}. This complexity is reflected 
in the size of the classical or quantum state space, which grows exponentially with the number of particles.
This exponential growth is reminiscent of the ``curse of dimensionality'' commonly encountered in machine learning.
That is, a target function to be learned requires an amount of training data that grows exponentially in the dimension (e.g.~the number of image features).
Despite this curse, the machine learning community has developed a number of techniques with remarkable
abilities to recognize, classify, and characterize complex sets of data. 
In light of this success, it is natural to ask whether such techniques could be applied to the arena of 
condensed-matter physics, particularly in cases where the microscopic Hamiltonian 
contains strong interactions, where numerical simulations are typically employed in the study of phases and phase transitions~\cite{avella2013strongly,Sandvik2010}. 
We demonstrate that modern machine learning architectures, such as fully-connected and convolutional neural networks~\cite{Goodfellow-et-al-2016-Book},
can provide a complementary approach to identifying phases and phase transitions in a variety of systems in condensed matter physics.
The training of neural networks on data sets obtained by Monte Carlo sampling provides a particularly 
powerful and simple framework for the supervised learning of phases and phase boundaries in physical models,
and can be easily built from readily-available tools such as Theano~\cite{bergstra+al:2010-scipy} or TensorFlow~\cite{tensorflow2015-whitepaper} libraries. 

Conventionally, the study of phases in condensed matter systems is performed with the help of tools that 
have been carefully designed to elucidate the underlying physical structures
of various states.
Among the most powerful are Monte Carlo simulations, which consist of two steps: a stochastic importance sampling over state space, and the evaluation of estimators 
for physical quantities calculated from these samples~\cite{Sandvik2010}. These estimators are constructed based on a variety of physical impetuses; e.g.~the ready availability 
of an analogous experimental measure like a specific heat; or, the encoding of some more abstract theoretical device, like an order parameter~\cite{wen2004quantum}.  
However, 
unique and technologically important states of matter may not be straightforwardly identified with standard 
estimators.  Indeed, for some highly-coveted phases such as topologically-ordered states~\cite{Kitaev20032,wen2004quantum}, positive identification may require prohibitively 
expensive (and experimentally challenging~\cite{Islam2015}) measures such as the entanglement entropy~\cite{LW,KP}.

\begin{figure}
\centering
\includegraphics[width=6.5in]{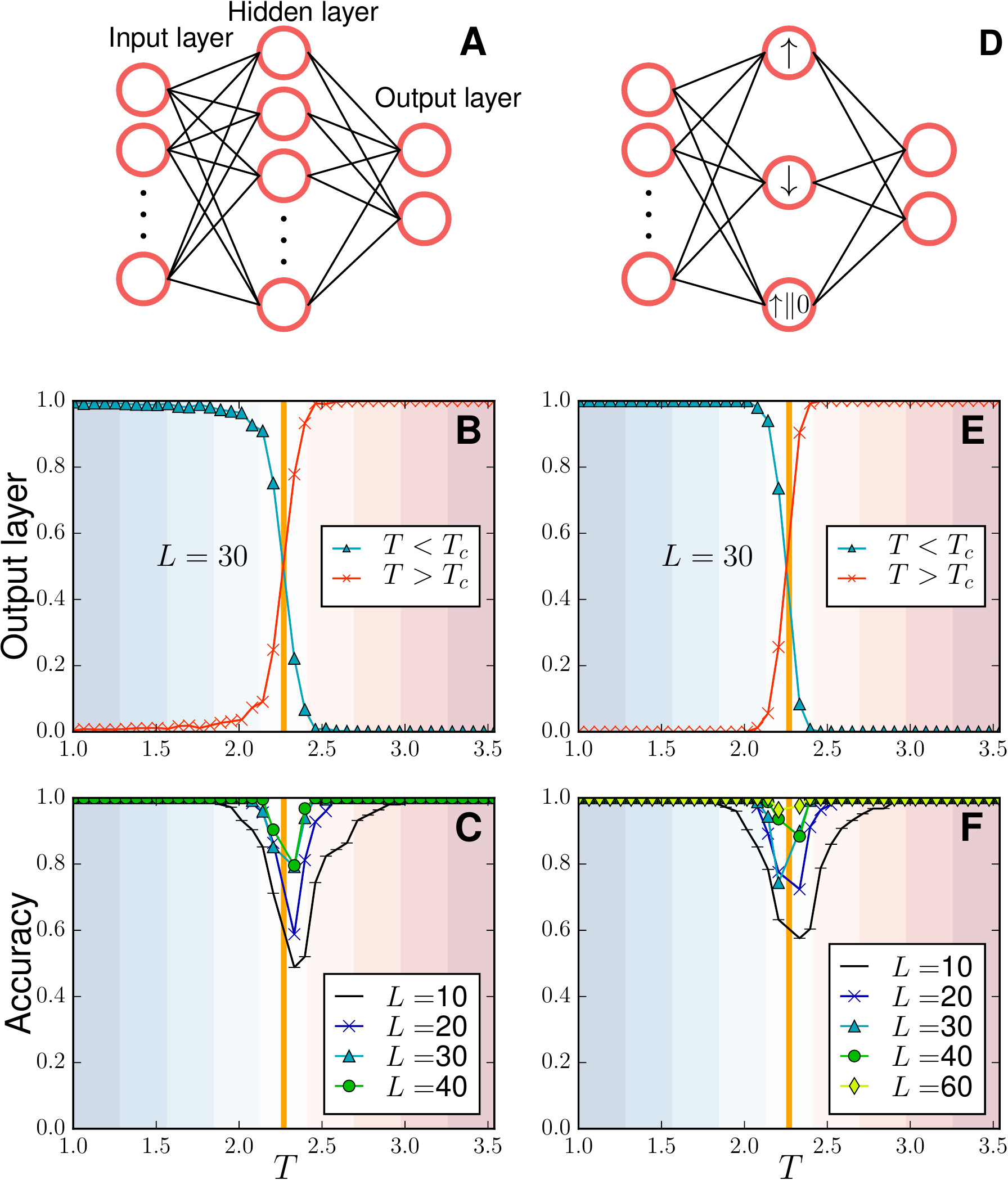}
\caption{Machine learning the ferromagnetic Ising model. (A) The trained neural network learns representations of 
the low- and high-temperature Ising states. The average output layer (B) and accuracy (C) over the test sets vs.~temperature. 
(D) Toy model of a neural network for the Ising model. (E) The average output layer and accuracy of the toy model 
are displayed in (E) and (F), respectively. The orange lines signal $T_c$ of the model in the thermodynamic limit, 
$T_c/J=2/\ln\left(1+\sqrt{2} \right)$~\cite{LarsyOnsy1944}. }
\label{fig:Isingneural}
\end{figure}

Machine learning, already explored as a tool in condensed matter and materials research
~\cite{Arsenault2014,Kusne2014,Kalinin2015,Ghiringhelli2015,Schoenholz2016,Mehta2014}, 
provides an alternate paradigm to the above approach.
The ability of modern machine learning techniques to classify, identify, or interpret 
massive data sets like images, videos, genome sequences, internet traffic statistics, natural language recordings, etc.~foreshadows their suitability to provide physicists 
with similar analyses on the exponentially large data sets embodied in the state space of condensed matter systems.  We first 
demonstrate this on
the prototypical example of the 
square-lattice ferromagnetic Ising model,  $H = - J\sum_{\langle i j \rangle}  \sigma^z_i \sigma^z_j$.
We set the energy scale $J=1$; the Ising variables $\sigma^z_i= \pm 1$ so that for $N$ lattice sites, the state space is of size $2^N$.
Standard Monte Carlo techniques can efficiently provide samples of configurations for any temperature $T$,
weighted by the Boltzmann distribution. The existence of a well-understood phase transition at temperature $T_c$~\cite{LarsyOnsy1944}, between a high-temperature 
paramagnetic phase and a low-temperature ferromagnetic phase, allows us the opportunity to attempt to classify the two different types of configurations 
without the use of Monte Carlo estimators (e.g. the magnetization).  Instead, we construct a fully connected feed-forward neural network,
implemented with TensorFlow~\cite{tensorflow2015-whitepaper}, to perform supervised learning directly on the raw configurations sampled by a Monte Carlo 
simulation (see Figure~\ref{fig:Isingneural}). The neural network is composed of an input layer with values determined by the spin configurations,  $100$-unit 
hidden layer of sigmoid neurons, and an analogous output layer. We use a cross-entropy cost function supplemented with an $L2$ regularization 
term to prevent overfitting. The neural network is trained using the Adam method for stochastic optimization~\cite{Kingma2014}. 
As Illustrated in Figure~\ref{fig:Isingneural}(A) through (C), when trained on a broad range of temperatures above and 
below $T_c$, the neural network is able to correctly classify $94\%$ of uncorrelated data provided in a test set, at the same temperatures as in the training set, 
for a system of $N=L^2=10^2$ spins. The classification accuracy improves as the system size is increased (as high as $99\%$ for $L=40$), as inferred from 
Figure~\ref{fig:Isingneural}(C), indicating that this training/testing paradigm is capable of systematically narrowing in on the true thermodynamic value of 
$T_c$ in a way analogous to the direct measurement of the magnetization in a conventional Monte Carlo simulation. 
In fact, due to the simplicity of the underlying order parameter (a bulk polarization of Ising spins below $T_c$), one can understand the training of the network 
through a simple toy model involving a hidden layer of only three analytically ``trained'' perceptrons, representing the possible combinations of high and 
low-temperature magnetic states exclusively based on their magnetization.
As illustrated in Figure~\ref{fig:Isingneural}(D) through (F), it performs the classification task with remarkably high accuracy. 
We emphasize that the toy model has no {\it a priori} knowledge of the critical temperature. Further details about the toy model, as well as a low-dimensional visualization of 
the training data to gain intuition for how these neural networks operate, are discussed in the supplementary materials. Similar results and success rates occur if the model 
is modified to have anti-ferromagnetic couplings, $H = J\sum_{\langle i j \rangle}  \sigma^z_i \sigma^z_j$, illustrating that the neural network is not only useful in 
identifying a global spin polarization, but an order parameter with a more complicated ordering wave vector ( here ${\bf q} = (\pi/a,\pi/a)$, where $a$ 
is the lattice spacing).  

\begin{figure}
\centering
\includegraphics[width=5.0in]{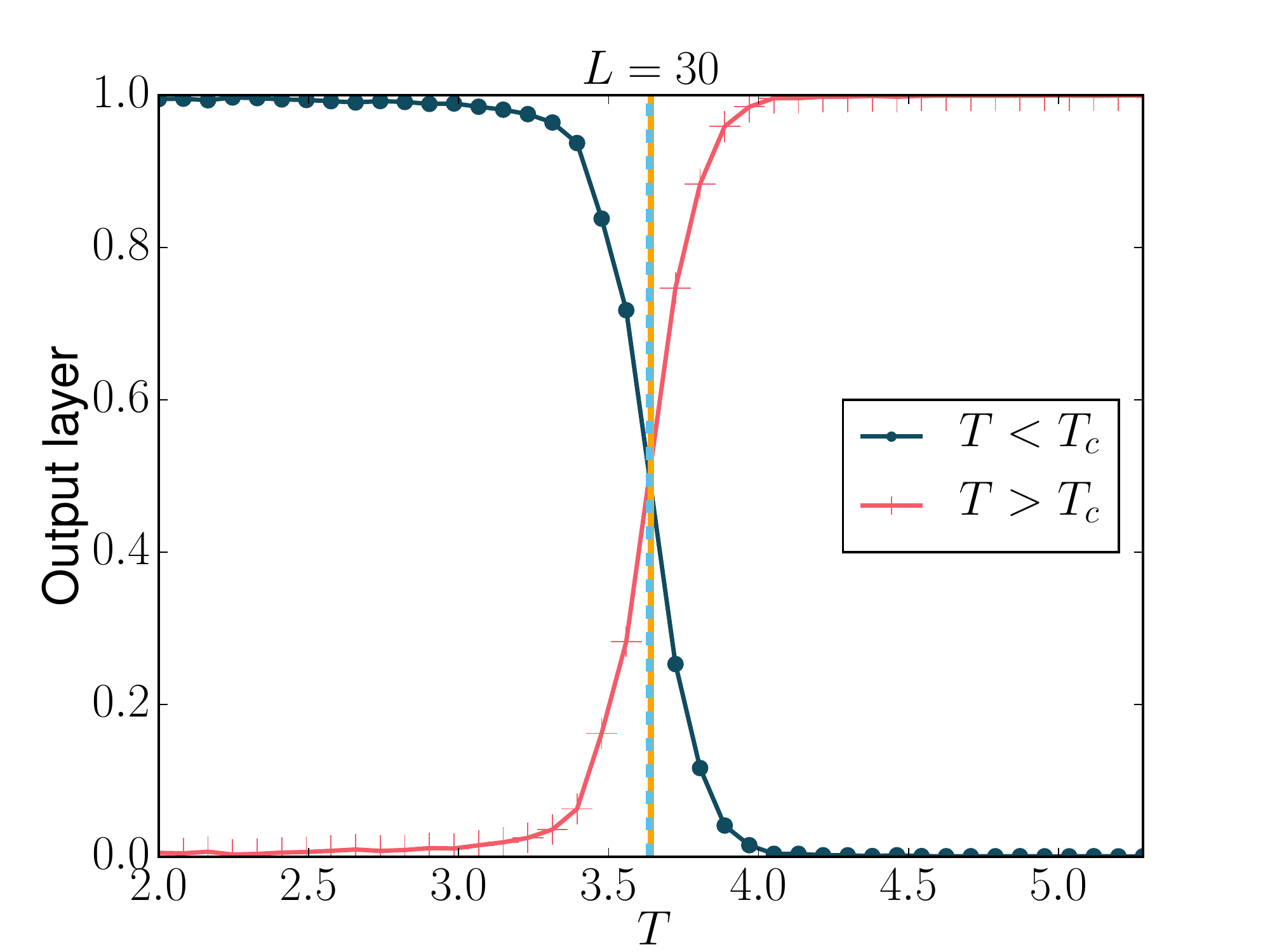}
\caption{Detecting the critical temperature of the triangular Ising model through the crossing of the values of the output layer vs $T$. 
The orange line signals the triangular Ising model $T_c/J=4/\ln3$, while the blue dashed line represents our estimate $T_c/J=3.63581$.}
\label{fig:triangle}
\end{figure}

Clearly, such a framework does not provide the same quantitive understanding as a direct Monte Carlo measurement of the order parameter, which sits on a 
solid bedrock of decades of statistical mechanics theory.  However, the power of neural networks
lies in their ability to generalize to tasks beyond their original design.  
For example, what if one was presented with a data set of Ising configurations from an unknown Hamiltonian, where the lattice structure (and therefore its $T_c$) is not known? 
We illustrate this scenario by taking our above feed-forward neural network, already trained on configurations for the square-lattice ferromagnetic Ising model, 
and feed it a test set produced by Monte Carlo simulations of the triangular lattice ferromagnetic Ising Hamiltonian. The network has no information about the Hamiltonian, the
lattice structure, or even the general locality of interactions. In Figure~\ref{fig:triangle} we present the output layer neurons averaged over the test set as a function 
of temperature for $L=30$. We estimate the critical temperature based on the crossing point of the low- and high-temperature outputs
to be $T_c/J=3.63581$, which is close to the exact thermodynamic $T_c/J =4/\ln 3\approx 3.640957$ \cite{Newell1950} -- a discrepancy easily attributed to finite-size effects.
Further, the same strategy can be repeated, using instead our toy neural network.  Again, without any knowledge of the critical temperatures on the square or triangular lattices, we estimate $T_c/J=3.63403$, differing from the true thermodynamic critical $T_c$ by less than $1\%$.  

\begin{figure}
\centering
\includegraphics[width=6.5in]{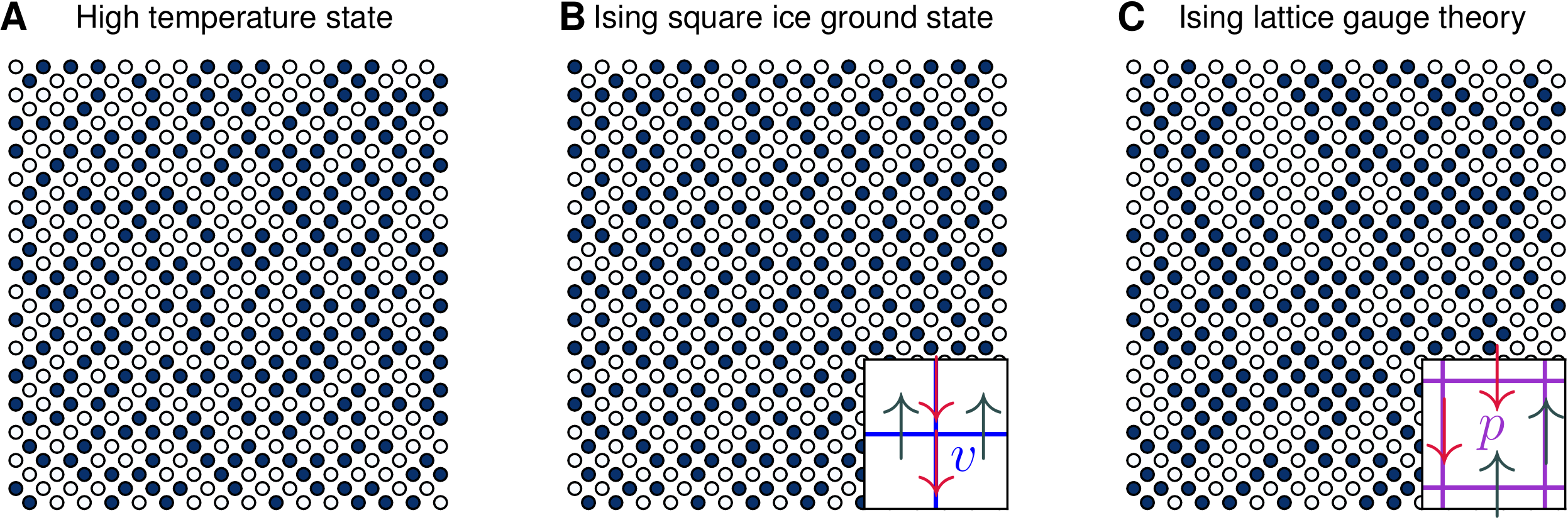}
\caption{Typical configurations of square ice and Ising gauge models. (A) A high-temperature state. (B) A ground state of the square ice Hamiltonian. 
(C) A ground state configuration of the Ising lattice gauge theory. The vertices and plaquettes defining the square ice and Ising gauge theory Hamiltonians are 
shown in the insets of (B) and (C).}
\label{fig:ice_tcode}
\end{figure}

We turn to the application of such techniques to problems of greater interest in modern condensed matter, such as disordered or topological phases, where no conventional order parameter exists. Coulomb phases, for example, are states of frustrated lattice models where local energetic constraints lead to extensively degenerate classical ground states,
which are highly-correlated ``spin liquids'' without a bulk magnetization or other local order parameter.
We consider a two-dimensional {\it square ice} Hamiltonian given by
$
H=J\sum_{v}Q_{v}^2
$
where the charge at vertex $v$ is $Q_v=\sum_{i\in v} \sigma_i^z$, and the Ising variables located in the lattice bonds as shown in Figure~\ref{fig:ice_tcode}. 
In a conventional condensed-matter approach, the ground states and the high-temperature states are distinguished by their spin-spin correlation functions:  
power-law decay in the Coulomb phase at $T=0$, and exponential decay at high temperature. Instead we use supervised learning, feeding raw Monte Carlo configurations 
to train a fully-connected neural network (Figure~\ref{fig:Isingneural}(A)) to distinguish ground states
from high-temperature states. Figure~\ref{fig:ice_tcode}(A) and Figure~\ref{fig:ice_tcode}(B) display high- and low-temperature 
snapshots of the configurations used in the training of the model. 
For a square ice system with $N=2\times16\times16$ spins,
we find that a standard fully-connected neural network with 100 hidden units successfully distinguishes the states with a $99\%$ accuracy.
The network does so solely based on spin configurations, with no information about the underlying lattice -- a feat difficult for the human eye, even if supplemented with 
a layout of the underlying Hamiltonian locality. 

These results indicate that the learning capabilities of neural networks go beyond the simple ability to encode order 
parameters, extending to the detection of subtle differences in higher-order correlations functions. 
As a final demonstration of this, we examine an Ising lattice gauge theory, one of the most 
prototypical examples of a topological phase of matter \cite{KogutRevModPhys.51.659,Kitaev20032}. The Hamiltonian is given by
$
H=-J\sum_{p}\prod_{i\in p}\sigma_i^z
$
where the Ising spins live on the bonds of a two-dimensional square lattice with plaquettes $p$, as 
shown in the inset of Figure~\ref{fig:ice_tcode}(C). The ground state is again a degenerate manifold \cite{Kitaev20032,Castelnovo2007} 
(Figure~\ref{fig:ice_tcode}(C)), with exponentially-decaying spin-spin correlations that makes it much more difficult to distinguish from the high temperature phase.

\begin{figure}
\centering
\includegraphics[width=6.5in]{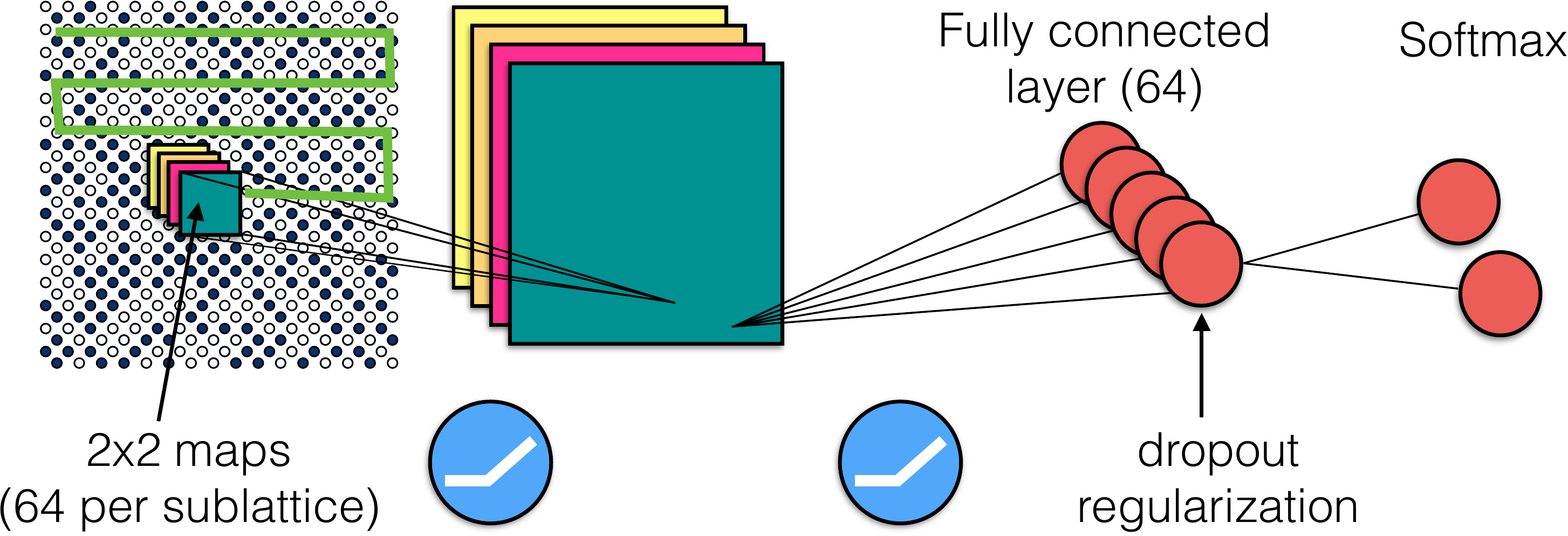}
\caption{Illustrating the convolutional neural network. The first hidden layer convolves 64  $2\times2$ filters with the spin configuration on each sublattice, followed 
by rectified linear units (ReLu). The outcome is followed by fully-connected layer with 64 units and a softmax output layer. The green line represents the sliding 
of the maps across the configuration.}
\label{fig:neural}
\end{figure}
Just as in the square ice model, we have made an attempt to use the neural network in Figure~\ref{fig:Isingneural}(A) to 
classify the high- and low- temperature states in the Ising gauge theory.
A straightforward implementation of supervised training fails to classify a test set containing samples of the two states to an 
accuracy over $50\%$ -- equivalent to simply guessing. Such failures typically occur because the neural network overfits to the training set. % \cite{Cybenko1989}.
To overcome this difficulty we consider a convolutional neural network 
(CNN)~\cite{Lecun98gradient-basedlearning,Goodfellow-et-al-2016-Book} which readily takes 
advantage of the two-dimensional structure of the input configurations, as well as the 
translational invariance of the model.
The CNN in Figure~\ref{fig:neural} is detailed in the supplementary materials. 
We optimize the CNN using Monte Carlo configurations drawn from the partition function of the 
Ising gauge theory at $T=0$ and $T=\infty$. Using this setting, the CNN successfully discriminates high-temperature from ground states 
with an accuracy of $100\%$ on a test set with $1\times10^{4}$ configurations, 
in spite of the lack of an order parameter or qualitative differences in the spin-spin correlations. 
Through the generation of new test sets that 
violate an extensive fraction of the local energetic constraints of the theory, we conclude that the discriminative 
power of the CNN relies on the detection of these satisfied constraints. 
Furthermore, test sets with defects that retain most local constraints but disrupt non-local features, 
like the extended closed-loop gas picture or the associated topological degeneracy \cite{Kitaev20032}, indicate that local constraints are the only features that the
CNN relies on for classification of the ground state. In view of these observations, we construct a simplified analytical toy model of our original CNN designed to 
explicitly exploit local constraints in the classification task. Such a model discriminates high-temperature from ground states with an accuracy of $100\%$. 
Details of the behavior of the CNN with various test sets, as well as the details of the analytical model, are contained in the supplementary material. 

We have shown that neural network technology, developed for engineering applications such as computer vision and natural language processing, can
be used to encode phases of matter and discriminate phase transitions in correlated many-body systems. In particular, we have argued that neural networks 
encode information about conventional ordered phases by learning the order parameter of the phase, 
without knowledge of the energy or locality conditions of Hamiltonian.
Furthermore, we have shown that neural networks can encode basic information about the ground states of unconventional disordered models, such as square ice model and the 
Ising lattice gauge theory, where they learn local constraints satisfied by the spin configurations in the
absence of an order parameter. These results indicate that neural networks have the potential to faithfully represent ground state wave functions. For instance,
ground states of the toric code~\cite{Kitaev20032,wen2004quantum} can be represented by convolutional neural networks akin to the one in Figure~\ref{fig:neural}  
(see the supplementary materials for details). 
We thus anticipate adoption to the field of quantum technology~\cite{Amin2016}, such as
quantum error correction protocols and quantum state tomography~\cite{Landon-Cardinal2012}.
The ability of machine learning algorithms to generalize to situations 
beyond their original design 
anticipates future applications such as the detection of phases and phase transitions in models vexed with 
the Monte Carlo sign problem~\cite{Sandvik2010}, as well as in experiments with single-site resolution capabilities such as the modern quantum gas 
microscopes~\cite{Bakr2009,Cheuk2015}. 
As in all other areas of ``big data'', we expect the rapid adoption of machine learning techniques 
as a basic research tool in condensed matter and statistical physics in the near future.

{\textit{Acknowledgments}}. 
We would like to thank Ganapathy Baskaran, Claudio Castelnovo, Anushya Chandran, Lauren E. Hayward Sierens, Bohdan Kulchytskyy, David Schwab, Miles Stoudenmire, Giacomo Torlai, 
Guifre Vidal, and Yuan Wan for discussions and encouragement. We thank Adrian Del Maestro for a careful reading of the manuscript. This research was supported 
by NSERC of Canada, the Perimeter Institute for Theoretical Physics, the John Templeton Foundation, and the Shared Hierarchical Academic Research Computing 
Network (SHARCNET). R.G.M.~acknowledges support from a Canada Research Chair. Research at Perimeter Institute is supported through Industry Canada and by the 
Province of Ontario through the Ministry of Research \& Innovation. 

\appendix

\section{Details of the toy model} \label{ToyModel}
\begin{figure}
\centering
\includegraphics[width=6.5in]{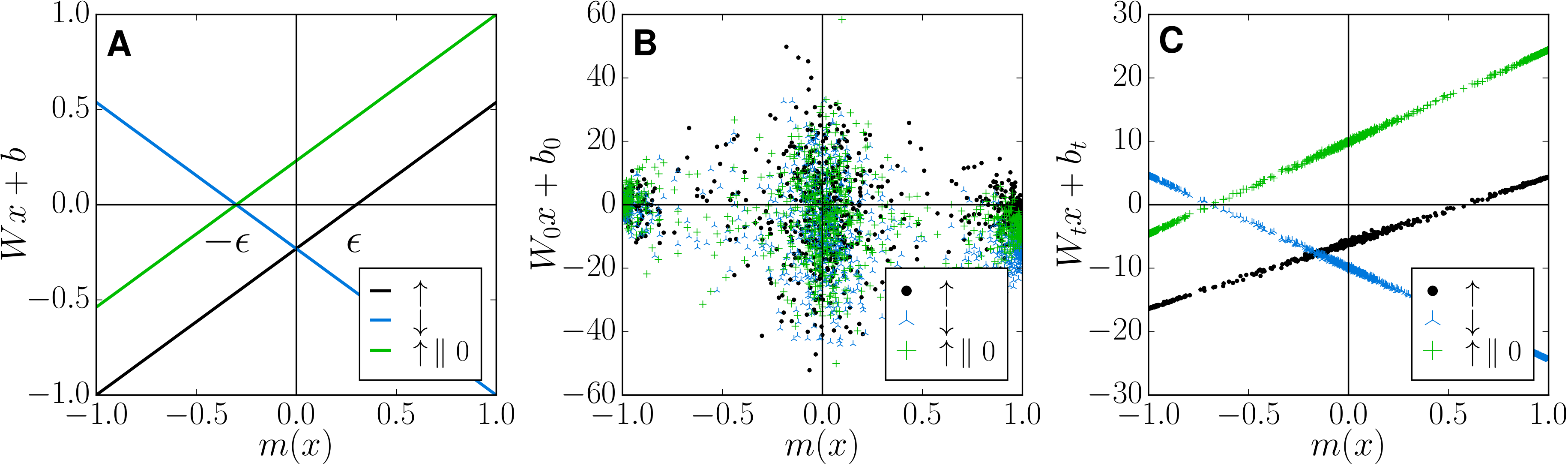}
\caption{Hidden layer arguments as a function of the magnetization of the Ising state $m(x)$. (A) Hidden layer arguments 
for our toy model. (B) and (C) show the arguments for a  neural network with 3 sigmoid neurons before and after training, respectively. 
%The training of the parameters, whose initial values are random, consists of $t=4.5\times10^4$ iterations and about 1687 epochs.
  } 
\label{fig:activations}
\end{figure}

The analytical model encodes the low- and high-temperature phases of the Ising model through their magnetization. The hidden layer contains 3 perceptrons (a neuron 
with a Heaviside step nonlinearity); the first two perceptrons activate when the input states are mostly polarized, while the third one activate if the 
states are polarized up or unpolarized. Notice that the third neuron can also be choosen to activate if the states are polarized down or unpolarized. The resulting 
outcomes are recombined in the output layer and produce the desired classification of the state. 
The hidden layer is parametrized through a weight matrix and bias vector given by
\begin{equation}
W= \frac{1}{N\left(1+\epsilon \right) } 
 \begin{pmatrix}
  1 & 1 & \cdots & 1 \\
  -1 & -1 & \cdots & -1 \\
   1 &  1 & \cdots & 1  
 \end{pmatrix}, \,\, \text{and} \,\,  
 b= \frac{\epsilon}{\left(1+\epsilon \right) }
 \begin{pmatrix}
  -1  \\
  -1 \\
   1 
 \end{pmatrix}, 
\end{equation}
where $0<\epsilon<1$ is the only free parameter of the model. The arguments of the three hidden layer neurons, in terms of  the weight matrix, bias vector, and 
a particular Ising configuration $x=\left(\sigma_1 \sigma_2, ...,\sigma_N\right)^{\text{T}}$, are given by
\begin{equation}
 Wx+b= \frac{1}{\left(1+\epsilon \right) }
 \begin{pmatrix}
   m(x)-\epsilon \\
  -m(x)-\epsilon \\
   m(x)+\epsilon 
 \end{pmatrix},
\end{equation}
where $m(x)=\frac{1}{N}\sum \limits_{i=1}^N \sigma_i$ is the magnetization of the Ising configuration. In Figure~\ref{fig:activations}(A) we display the components of the $Wx+b$ vector as a function of the
magnetization of the Ising state $m(x)$. The first and second neuron activate when the state is predominantly polarized, i.e., when $m(x)>\epsilon$ or $m(x)<-\epsilon$.
The third neuron activates if the state has a magnetization $m(x)>-\epsilon$, which means that, in the limit where $0<\epsilon\ll 1$, it activates when the state is 
either polarized or unpolarized. The parameter $\epsilon$ is thus a threshold value of the magnetization that helps deciding whether the state is considered polarized 
or not. %For large system sizes, the optimal value the threshold is typically $\epsilon\approx0.65$.

The output layer is parametrized through a weight matrix and bias vector given by
\begin{equation}
W_2= 
 \begin{pmatrix}
  2 & 1 & -1 \\
  -2 & -2 & 1 
 \end{pmatrix}, \,\, \text{and} \,\,
 b_2= 
 \begin{pmatrix}
  0  \\
  0 
 \end{pmatrix},
\end{equation}

where these arbitrary choices ensure that the ordered, low-$T$ output neuron $O_{\text{Low-T}}=1$ is active when either the spins polarize mostly $\uparrow$ or 
$\downarrow$. On the other hand, when the $\uparrow \,\parallel 0 $ neuron is active but the $\uparrow$ is not, then the high-temperature output 
neuron $O_{\text{High-T}}=1$, symbolizing a high-temperature state.

To illustrate what the effects of the training on the parameters $W$ and $b$ are, we consider the numerical training of a fully-connected neural network 
with only 3 neurons using the same setup and training/test data used in our ferromagnetic model (Figure~\ref{fig:Isingneural}) 
for the $L=30$ system. In Figure~\ref{fig:activations}(B) we display the argument $W_0x+b_0$ of the input layer at training iteration $t=0$ for 
configurations $x$ in the test set as a function of the magnetization of the configurations $m(x)$. The weights have been randomly initialized at $t=0$ 
from a normal distribution with zero mean and unit standard deviation. As the training proceeds, the parameters adjust such that the components of the  
vector $W_tx+b_t$ approximately become linear functions of the magnetization $m(x)$, as shown in Figure~\ref{fig:activations}(C), in agreement with the 
assumptions of our toy model. These results clearly support our claim about the neural network's ability encode and {\it learn} the magnetization in the hidden layer.  

\section{Visualizing the action of a neural network on the Ising ferromagnet}
\begin{figure}
\centering
\includegraphics[width=4.0in]{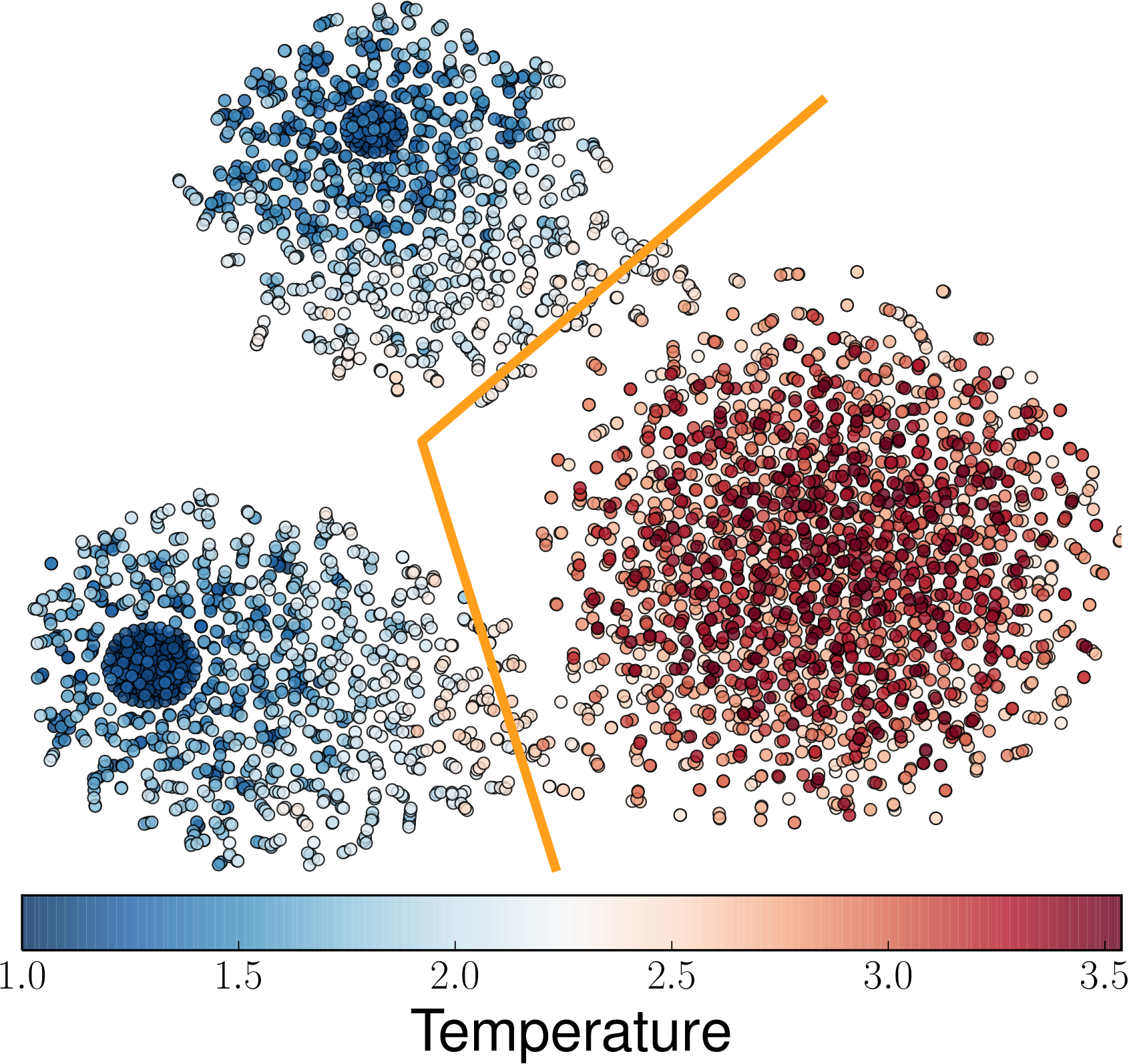}
\caption{Two-dimensional t-SNE visualization of the training set used in the Ising model for $L=30$ colored according to temperature.
The orange line represents a hyperplane separating the low- from high-temperatures states. }
\label{fig:tsneIsing}
\end{figure}

A strategy to gain intuition for how these neural networks operate is to produce a low-dimensional visualization of data used in the training. We consider the
t-distributed stochastic neighbor embedding (t-SNE) technique~\cite{t-SNE} where high-dimensional data is embedded in two or three dimensions so
that data points close to each other in the original space are also positioned close to each other in the embedded low-dimensional space. Figure~\ref{fig:tsneIsing}
displays a t-SNE visualization of the Ising configurations used the training of our ferromagnetic model. The two low-temperature blue regions correspond to
the two ordered states with spins polarized either up or down. The high-temperature red region identifies the paramagnetic state. The resulting neural networks,
which are functions defined over the high-dimensional state space, become trained so that the low-temperature output neuron takes a high value in the cool region
(and vice versa), crossing over to a low value as the system is warmed through the orange hyperplane. This allows the classification of a state in terms of the
neuron values.

\section{Details of the convolutional neural network of the Ising lattice gauge theory}

The exact architecture of the convolutional neural network (CNN)~\cite{Goodfellow-et-al-2016-Book}, schematically described in Figure~\ref{fig:neural}, is as follows. 
The input layer is a two-dimensional Ising spin configuration with $N=16\times16\times2$ spins, where $\sigma_i=\pm1$. The first hidden layer convolves 64 $2\times 2$ 
filters on each of the two sublattices of the model with a unit stride, no padding, with periodic boundary conditions, followed by rectified linear unit (ReLu). The 
final hidden layer is a fully-connected layer with 64 ReLu units, while the output is a softmax layer with two outputs (correponding to $T=0$ and $T=\infty$ states). 
To prevent overfitting, we apply a dropout regularization in the fully-connected layer~\cite{dropout}. Our model has been implemented using 
TensorFlow~\cite{tensorflow2015-whitepaper}.  

\begin{figure}
\centering
\includegraphics[width=6.5in]{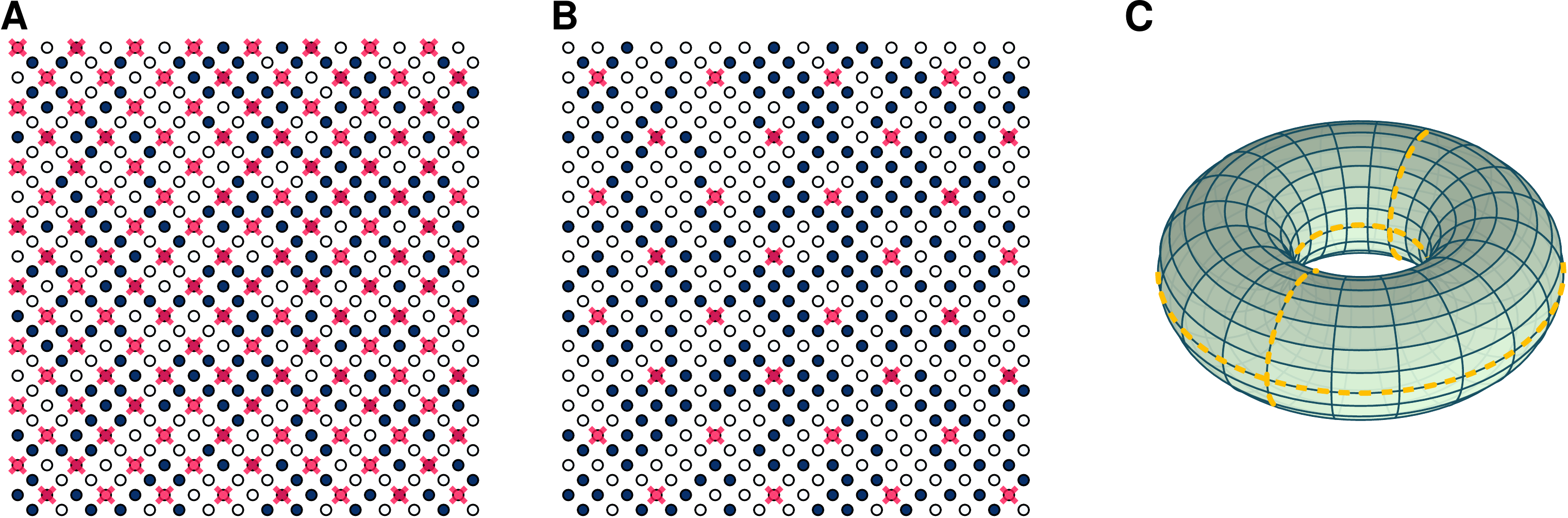}
\caption{Investigating the source of discriminative power of the convolutional neural network. New test sets with sublattice rotations where the local constrains 
are such that $1/2$ spins per plaquette (A) and $1/8$ spins per plaquette (B) are flipped. The red crossess symbolize the flipped spins. (C) Cuts/stitches 
(yellow dashed lines) performed on the ground state configurations in order to produce a new test set from mixing the 4 resulting pieces among different ground states.}
\label{fig:toruscuts}
\end{figure}

Since our CCN correctly classifies $T=0$ and $T=\infty$ states with $100\%$ accuracy, we would like to scrutinize the origin of its discriminiative 
power by asking whether it discerns the states by the presence (or absence) of the local Hamiltonian constraints or the extended closed-loop structure. 
Our strategy consists in the construction of new test sets with modified low temperature states, as detailed below. First, we consider transformations
that do not destroy the topological order~\cite{Castelnovo2007} of the $T=0$ state but change the local constraints of the original Ising lattice
 gauge theory. As shown in the configurations in Figure~\ref{fig:toruscuts}(A) and (B), we consider transformations where a spin is flipped every 
$m=2$ (A) ($m=8$ (B)) plaquettes. The positions of the flipped spins are marked with red crosses.  After optimizing the CNN using the original training set, 
the neural network classifies most of the transformed $T=0$ states as high-temperature ones, resulting in an overall test accuracy of $50\%$ and $55\%$ for $m=2$ 
and $m=8$, respectively. This reveals that the neural network relies on the presence of satisfied local constraints of the original Ising lattice gauge theory, and not
on the topological order of the state, in deciding whether a state is considered low or high temperature. Second, we consider a new test set where the $T=0$ 
states retain most local constraints but disrupt non-local features like the extended closed-loop structure. We consider dividing the original states into 4 pieces 
as shown in Figure~\ref{fig:toruscuts}(C) and then reshuffling the 4 pieces among different states, subsequently stitching them to form new ``low'' temperature 
configurations. The new configurations will contain defects along the dashed lines in Figure~\ref{fig:toruscuts}(C), thus disrupting the extended closed-loop picture,  
but preserving the local constraints everywhere else in the configuration. We find that the trained CNN recognizes such states as ground states with high confidence, 
suggesting that the CNN does not use the extended closed-loop structure and indicating that local constraints are the only features that the
CNN relies on for classification of the ground state. 

In view of the conclusion above, we now present a toy model that uses a streamlined version of our original CNN constructed to explicitly
detect satisfied energetic local constraints. The convolutional layer contains 16 2$\times$2 filters per sublattice with unit stride in both directions 
and periodic boundary conditions. The convolutional layer is fully connected to two perceptron neurons in the output layer, as described below.
\begin{figure}
\centering
\includegraphics[width=4.0in]{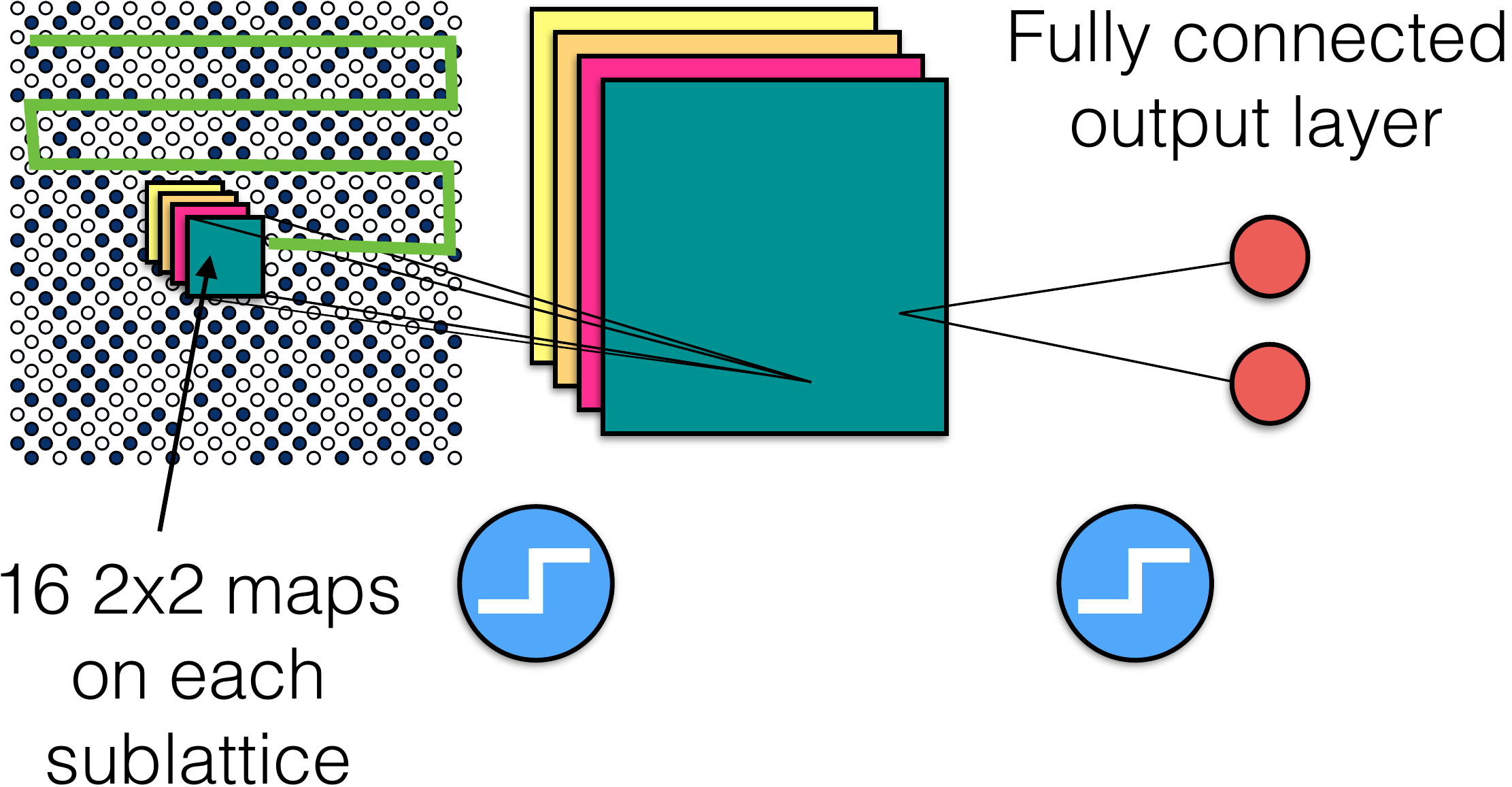}
\caption{Toy model of a convolutional neural network to classify states of the Ising gauge theory.}
\label{fig:toyIsingGauge}
\end{figure}
A schematic representation of the toy CNN is presented in Figure~\ref{fig:toyIsingGauge}. The values of the filters $W_{y x s f }$ are presented in 
Table ~\ref{W-table}, where $x$ and $y$ represent the spatial indices of the convolution, and $s$ and $f$ label the sublattice and the filter, respectively. 
The purpose of the filters is to individually process each plaquette in the spin configuration and determine whether its energetic constraints are satisfied or not. 
The Ising gauge theory contains $16$ different spin configurations per plaquette, of which 8 satisfy the energetic constraints of the Hamiltonian. The first 
group of 16 filters  $W_{y x s f }$ ($f=1$ thorugh $f=8$, left blue column in Table ~\ref{W-table}) detect satisfied plaquettes, while the remaining 16 ($f=9$ through $f=16$, 
right red column in Table ~\ref{W-table}) detect unsatisfied plaquettes. The bias of the convolutional layer is a 16-dimensional vector given by 
$b_{c}=-(2+\delta)\left( 1 \cdots 1 \right)^{\text{T}}$, where $0<\delta\ll1$ is a small parameter. The outcome of the convolutional layer consists of 16 two-dimensional 
arrays of size $L \times L$ processed through perceptrons. Here the total number of spins in the configuration is $N=2\times L\times L$, where $L$ is the linear 
size of the system.  The outcome of the convolutional layer is reshaped into a $16 \times L^2$-dimensional vector such that the first $8L^2$ entries correspond 
to the outcome of the first group of filters ($f=1$ thorugh $f=8$) while the remaining $8L^2$ correspond to the last group of filters ($f=9$ through $f=16$). The output layer,
which is fully connected to the the convolutional layer, contains two perceptron neurons denoted by $O_{0}$ and $O_{\infty}$ for zero- and high-temperature states, 
respectively. It is parametrized through a weight matrix and a bias vector given by 
\begin{table}[]
\centering
\label{W-table}
\begin{tabular}{|>{\columncolor[RGB]{164,222,249}[.8\tabcolsep]}l|l|l||>{\columncolor[RGB]{253,197,245}[.8\tabcolsep]}l|l|l|}
\hline
f & s=A & s=B & f & s=A & s=B \\ \hline
1 \,  & $\scriptscriptstyle \begin{pmatrix*}[r]1& 0\\[-0.8em] 1&0\end{pmatrix*}$ & $\scriptscriptstyle \begin{pmatrix*}[r] 1&1 \\[-0.8em] 0& 0\end{pmatrix*}$ & 
9  & $\scriptscriptstyle \begin{pmatrix*}[r] 1& 0\\[-0.8em] 1&0\end{pmatrix*}$ & $\scriptscriptstyle \begin{pmatrix*}[r] 1&-1 \\[-0.8em] 0& 0\end{pmatrix*}$ \\ \hline 
2 \, & $\scriptscriptstyle \begin{pmatrix*}[r]-1& 0\\[-0.8em]-1&0\end{pmatrix*}$ & $\scriptscriptstyle \begin{pmatrix*}[r]-1&-1 \\[-0.8em] 0& 0\end{pmatrix*}$ &
10 & $\scriptscriptstyle \begin{pmatrix*}[r] 1& 0\\[-0.8em] 1&0\end{pmatrix*}$ & $\scriptscriptstyle \begin{pmatrix*}[r]-1& 1 \\[-0.8em] 0& 0\end{pmatrix*}$ \\ \hline
3 \, & $\scriptscriptstyle \begin{pmatrix*}[r] 1& 0\\[-0.8em] 1&0\end{pmatrix*}$ & $\scriptscriptstyle \begin{pmatrix*}[r]-1&-1 \\[-0.8em] 0& 0\end{pmatrix*}$ & 
11 & $\scriptscriptstyle \begin{pmatrix*}[r] 1& 0\\[-0.8em]-1&0\end{pmatrix*}$ & $\scriptscriptstyle \begin{pmatrix*}[r] 1& 1 \\[-0.8em] 0& 0\end{pmatrix*}$ \\ \hline
4  \,& $\scriptscriptstyle \begin{pmatrix*}[r] 1& 0\\[-0.8em]-1&0\end{pmatrix*}$ & $\scriptscriptstyle \begin{pmatrix*}[r]-1& 1 \\[-0.8em] 0& 0\end{pmatrix*}$ & 
12 & $\scriptscriptstyle \begin{pmatrix*}[r]-1& 0\\[-0.8em] 1&0\end{pmatrix*}$ & $\scriptscriptstyle \begin{pmatrix*}[r] 1& 1 \\[-0.8em] 0& 0\end{pmatrix*}$ \\ \hline
5  \,& $\scriptscriptstyle \begin{pmatrix*}[r]-1& 0\\[-0.8em]-1&0\end{pmatrix*}$ & $\scriptscriptstyle \begin{pmatrix*}[r] 1& 1 \\[-0.8em] 0& 0\end{pmatrix*}$ &    
13 & $\scriptscriptstyle \begin{pmatrix*}[r]-1& 0\\[-0.8em]-1&0\end{pmatrix*}$ & $\scriptscriptstyle \begin{pmatrix*}[r]-1& 1 \\[-0.8em] 0& 0\end{pmatrix*}$ \\ \hline
6  \,& $\scriptscriptstyle \begin{pmatrix*}[r]-1& 0\\[-0.8em] 1&0\end{pmatrix*}$ & $\scriptscriptstyle \begin{pmatrix*}[r] 1&-1 \\[-0.8em] 0& 0\end{pmatrix*}$ & 
14 & $\scriptscriptstyle \begin{pmatrix*}[r]-1& 0\\[-0.8em]-1&0\end{pmatrix*}$ & $\scriptscriptstyle \begin{pmatrix*}[r] 1&-1 \\[-0.8em] 0& 0\end{pmatrix*}$ \\ \hline
7  \,& $\scriptscriptstyle \begin{pmatrix*}[r] 1& 0\\[-0.8em]-1&0\end{pmatrix*}$ & $\scriptscriptstyle \begin{pmatrix*}[r] 1&-1 \\[-0.8em] 0& 0\end{pmatrix*}$ &   
15 & $\scriptscriptstyle \begin{pmatrix*}[r]-1& 0\\[-0.8em] 1&0\end{pmatrix*}$ & $\scriptscriptstyle \begin{pmatrix*}[r]-1&-1 \\[-0.8em] 0& 0\end{pmatrix*}$ \\ \hline
8  \,& $\scriptscriptstyle \begin{pmatrix*}[r]-1& 0\\[-0.8em] 1&0\end{pmatrix*}$ & $\scriptscriptstyle \begin{pmatrix*}[r]-1& 1 \\[-0.8em] 0& 0\end{pmatrix*}$ & 
16 & $\scriptscriptstyle \begin{pmatrix*}[r]-1& 0\\[-0.8em]-1&0\end{pmatrix*}$ & $\scriptscriptstyle \begin{pmatrix*}[r]-1&-1 \\[-0.8em] 0& 0\end{pmatrix*}$    \\ \hline
\end{tabular}
\caption{Specifying the tensor $W_{yxsf}$ filters in the CNN. $y$ and $x$ specify the spatial indices of the filter, while $s$ and $f$ specify the sublattice and the filter, respectively. }
\end{table}
\begin{equation}
W_{\text{o}}=\left( \begin{matrix} \overbrace{ \, 1 \,\,\, \hdots \,\,\, 1}^{\scriptscriptstyle \, 8L^2 \, \text{terms}} & \overbrace{-L^2+L-1 \hdots -L^2+L-1 }^{\scriptscriptstyle \, 8L^2\, \text{terms}}\\ -\!1 \,\, \hdots \,-\!\!1  & L^2-L+1\hdots L^2-L+1 \end{matrix}\right), \,\,\,\, \text{and} \,\, b_{\text o}=\left(\begin{matrix}0\\0\end{matrix}\right)
\end{equation}
These choices ensure that whenever an unsatisfied plaquette is encountered by the convolutional layer, the zero-temperature neuron is $O_{0}=0$ and the high-temperature 
$O_{\infty}=1$, while only if {\it all} energetic constraints are satisfied $O_{0}=1$ and $O_{\infty}=0$, thus allowing the classification of the states. When used
on our test sets, the model performs the classification task with a $100\%$ accuracy, which means that all the high temperature states in the test set contain least 
one unsatisfied plaquette. Note that the classification error for this task is expected to be exponentially small in the volume of the system, since at infinte temperature the 
ground states appear with exponentially small probability. Having distilled the model's basic ingredients, we proceed to train an analogue model {\it numerically} 
starting from random weights and biases $W_{yxsf}$, $W_{\text{o}}$, $b_{\text{c}}$, and $b_{\text{o}}$. Further, we replace the perceptron nonlinearities by ReLu units and
a softmax output layer to enable a reliable numerical training. After the training, the model performs the classification task with a $100\%$ accuracy on the test sets, 
as expected.

As a consequence of the classification scheme provided by the analytical toy model, we observe that the values of the zero-temperature neuron $O_{0}$ behave exactly like 
the amplitudes of one of the ground states of the toric code written in the $\sigma_z$ basis~\cite{Kitaev20032}. The ground state described by $O_{0}$ is a linear 
combination of all 4 ground states with well defined parity on the torus. More precisely, such a state can be written as  
$|\psi_{\text{toric}} \rangle=\sum_{\sigma_{z1},...,\sigma_{zN}}O_{0}(\sigma_{z1}...\sigma_{z N}) |\sigma_{z 1}...\sigma_{z N}\rangle $, 
where the spin configurations $\sigma_{zi}=\pm1$, and $O_{0}(\sigma_{z 1}...\sigma_{z N})$ corresponds to the value of $O_{0}$ 
after a feed-forward pass of the neural network for a given a input configuration $\sigma_{z\, 1},...,\sigma_{z\, N}$. Our model bears resemblance with the 
construction of the ground state of the toric code in terms of projected entangled pair states in that local tensors project out states 
containing plaquettes with odd parity~\cite{PhysRevLett.109.260401}. These observations suggest that convolutional neural networks have the potential to represent 
ground states with topological order.

%\bibliographystyle{Science}
%\bibliography{scibib}

\end{document}